\documentclass[twocolumn,prl,showpacs,preprintnumbers,amsmath,amssymb,superscriptaddress,nofootinbib]{revtex4-2}

\usepackage{graphicx}% Include figure files
\usepackage{dcolumn}% Align table columns on decimal point
\usepackage{bm}% bold math
\usepackage{booktabs}
\usepackage{placeins}
\usepackage{rotating}
\usepackage{color}
\usepackage[mathlines]{lineno}% Enable numbering of text and display math
\begin{document}

%\preprint{To be submitted to Physical Review Letters}
%\preprint{APS/123-QED}

\title{Mapping the \boldmath $N = 40$ Island of Inversion:
Precision Mass Measurements of Neutron-rich Fe Isotopes}% Force line breaks with \\
\author{W. S. Porter}
\email[Corresponding author: ]{wporter@triumf.ca}
    \affiliation{TRIUMF, 4004 Wesbrook Mall, Vancouver, British Columbia V6T 2A3, Canada}
    \affiliation{Department of Physics \& Astronomy, University of British Columbia, Vancouver, British Columbia V6T 1Z1, Canada} 
    
\author{B. Ashrafkhani}
    \affiliation{Department of Physics \& Astronomy, University of Calgary, Calgary, Alberta T2N 1N4, Canada}
    
\author{J. Bergmann}
    \affiliation{II. Physikalisches Institut, Justus-Liebig-Universit\"{a}t, 35392 Gie{\ss}en, Germany}
    
\author{C. Brown}
    \affiliation{School of Physics and Astronomy, University of Edinburgh, Edinburgh, EH9 3FD, United Kingdom}

\author{T. Brunner}
    \affiliation{TRIUMF, 4004 Wesbrook Mall, Vancouver, British Columbia V6T 2A3, Canada}
    \affiliation{Department of Physics, McGill University, 3600 Rue University, Montr\'eal, QC H3A 2T8, Canada}
    
\author{J. D. Cardona}
    \affiliation{TRIUMF, 4004 Wesbrook Mall, Vancouver, British Columbia V6T 2A3, Canada}
    \affiliation{Department of Physics \& Astronomy, University of Manitoba, Winnipeg, Manitoba R3T 2N2, Canada}

\author{D. Curien}
    \affiliation{Universit\'{e} de Strasbourg, CNRS, IPHC UMR 7178, F-67 037 Strasbourg, France}
    
\author{I. Dedes}
    \affiliation{Institute of Nuclear Physics, Polish Academy of Sciences, PL-31 342 Krak\'{o}w, Poland}
    
\author{T. Dickel}
    \affiliation{II. Physikalisches Institut, Justus-Liebig-Universit\"{a}t, 35392 Gie{\ss}en, Germany}
    \affiliation{GSI Helmholtzzentrum f\"{u}r Schwerionenforschung GmbH, Planckstra{\ss}e 1, 64291 Darmstadt, Germany}

\author{J. Dudek}
    \affiliation{Universit\'{e} de Strasbourg, CNRS, IPHC UMR 7178, F-67 037 Strasbourg, France}
    \affiliation{Institute of Physics, Marie Curie-Sk\l{}odowska University, PL-20 031 Lublin, Poland}  
\author{E. Dunling}
    \affiliation{TRIUMF, 4004 Wesbrook Mall, Vancouver, British Columbia V6T 2A3, Canada}
    \affiliation{Department of Physics, University of York, York, YO10 5DD, United Kingdom}

\author{G. Gwinner}
    \affiliation{Department of Physics \& Astronomy, University of Manitoba, Winnipeg, Manitoba R3T 2N2, Canada}
    
\author{Z. Hockenbery}
    \affiliation{TRIUMF, 4004 Wesbrook Mall, Vancouver, British Columbia V6T 2A3, Canada}
    \affiliation{Department of Physics, McGill University, 3600 Rue University, Montr\'eal, QC H3A 2T8, Canada}
    
\author{J. D. Holt}
    \affiliation{TRIUMF, 4004 Wesbrook Mall, Vancouver, British Columbia V6T 2A3, Canada}    
    \affiliation{Department of Physics, McGill University, Montr\'eal, QC H3A 2T8, Canada}
    
\author{C. Hornung}
    \affiliation{GSI Helmholtzzentrum f\"{u}r Schwerionenforschung GmbH, Planckstra{\ss}e 1, 64291 Darmstadt, Germany}
    
\author{C. Izzo}
    \affiliation{TRIUMF, 4004 Wesbrook Mall, Vancouver, British Columbia V6T 2A3, Canada}

\author{A. Jacobs}
    \affiliation{TRIUMF, 4004 Wesbrook Mall, Vancouver, British Columbia V6T 2A3, Canada}
    \affiliation{Department of Physics \& Astronomy, University of British Columbia, Vancouver, British Columbia V6T 1Z1, Canada}
    
\author{A. Javaji}
    \affiliation{TRIUMF, 4004 Wesbrook Mall, Vancouver, British Columbia V6T 2A3, Canada}
    \affiliation{Department of Physics \& Astronomy, University of British Columbia, Vancouver, British Columbia V6T 1Z1, Canada} 
    
\author{B. Kootte}
    \affiliation{TRIUMF, 4004 Wesbrook Mall, Vancouver, British Columbia V6T 2A3, Canada}
    \affiliation{Department of Physics \& Astronomy, University of Manitoba, Winnipeg, Manitoba R3T 2N2, Canada}

\author{G. Kripk\'{o}-Koncz}
    \affiliation{II. Physikalisches Institut, Justus-Liebig-Universit\"{a}t, 35392 Gie{\ss}en, Germany}

\author{E. M. Lykiardopoulou}
    \affiliation{TRIUMF, 4004 Wesbrook Mall, Vancouver, British Columbia V6T 2A3, Canada}
    \affiliation{Department of Physics \& Astronomy, University of British Columbia, Vancouver, British Columbia V6T 1Z1, Canada}
    
\author{T. Miyagi}
    \affiliation{TRIUMF, 4004 Wesbrook Mall, Vancouver, British Columbia V6T 2A3, Canada}   
    
\author{I. Mukul}
    \affiliation{TRIUMF, 4004 Wesbrook Mall, Vancouver, British Columbia V6T 2A3, Canada}
    
\author{T. Murb\"{o}ck}
    \affiliation{TRIUMF, 4004 Wesbrook Mall, Vancouver, British Columbia V6T 2A3, Canada}
    
\author{W. R. Pla\ss}
    \affiliation{II. Physikalisches Institut, Justus-Liebig-Universit\"{a}t, 35392 Gie{\ss}en, Germany}
    \affiliation{GSI Helmholtzzentrum f\"{u}r Schwerionenforschung GmbH, Planckstra{\ss}e 1, 64291 Darmstadt, Germany}
    
\author{M. P. Reiter}
    \affiliation{TRIUMF, 4004 Wesbrook Mall, Vancouver, British Columbia V6T 2A3, Canada}
    \affiliation{II. Physikalisches Institut, Justus-Liebig-Universit\"{a}t, 35392 Gie{\ss}en, Germany}
    \affiliation{School of Physics and Astronomy, University of Edinburgh, Edinburgh, EH9 3FD, United Kingdom}
    
\author{J. Ringuette}
    \affiliation{TRIUMF, 4004 Wesbrook Mall, Vancouver, British Columbia V6T 2A3, Canada}
    \affiliation{Department of Physics, Colorado School of Mines, Golden, Colorado 80401, USA}

\author{C. Scheidenberger}
    \affiliation{II. Physikalisches Institut, Justus-Liebig-Universit\"{a}t, 35392 Gie{\ss}en, Germany}
    \affiliation{GSI Helmholtzzentrum f\"{u}r Schwerionenforschung GmbH, Planckstra{\ss}e 1, 64291 Darmstadt, Germany}
    \affiliation{Helmholtz Forschungsakademie Hessen f\"{u}r FAIR (HFHF), GSI Helmholtzzentrum f\"{u}r Schwerionenforschung, Campus Gie{\ss}en, 35392 Gie{\ss}en, Germany}
    
\author{R. Silwal}
    \altaffiliation{Current address: Department of Physics and Astronomy, Appalachian State University, Boone, North Carolina 28608, USA}
    \affiliation{TRIUMF, 4004 Wesbrook Mall, Vancouver, British Columbia V6T 2A3, Canada}
    
\author{C. Walls}
    \affiliation{TRIUMF, 4004 Wesbrook Mall, Vancouver, British Columbia V6T 2A3, Canada}
    \affiliation{Department of Physics \& Astronomy, University of Manitoba, Winnipeg, Manitoba R3T 2N2, Canada}

\author{H. L. Wang}
    \affiliation{School of Physics and Microelectronics, Zhengzhou University, Zhengzhou 4500001, China}
    
\author{Y. Wang}
    \affiliation{TRIUMF, 4004 Wesbrook Mall, Vancouver, British Columbia V6T 2A3, Canada}
    \affiliation{Department of Physics \& Astronomy, University of British Columbia, Vancouver, British Columbia V6T 1Z1, Canada}

\author{J. Yang}
    \affiliation{Institute of Physics, Marie Curie-Sk\l{}odowska University, PL-20 031 Lublin, Poland}
    \affiliation{School of Physics and Microelectronics, Zhengzhou University, Zhengzhou 4500001, China}
    
\author{J. Dilling}
    \affiliation{TRIUMF, 4004 Wesbrook Mall, Vancouver, British Columbia V6T 2A3, Canada}
    \affiliation{Department of Physics \& Astronomy, University of British Columbia, Vancouver, British Columbia V6T 1Z1, Canada}
    
\author{A. A. Kwiatkowski}
    \affiliation{TRIUMF, 4004 Wesbrook Mall, Vancouver, British Columbia V6T 2A3, Canada}
    \affiliation{Department of Physics and Astronomy, University of Victoria, Victoria, British Columbia V8P 5C2, Canada}
    
%\affiliation{
%\affiliation{
%\affiliation{%

\date{\today}% It is always \today, today,

\begin{abstract}
Nuclear properties across the chart of nuclides are key to improving and validating our understanding of the strong interaction in nuclear physics. We present high-precision mass measurements of neutron-rich Fe isotopes performed at the TITAN facility. The multiple-reflection time-of-flight mass spectrometer (MR-ToF-MS), achieving a resolving power greater than $600\,000$ for the first time, enabled the measurement of $^{63-70}$Fe, including first-time high-precision direct measurements ($\delta m/m \sim 10^{-7}$) of $^{68-70}$Fe, as well as the discovery of a long-lived isomeric state in $^{69}$Fe. These measurements are accompanied by both mean-field and ab initio calculations using the most recent realizations which enable theoretical assignment of the spin-parities of the $^{69}$Fe ground and isomeric states. Together with mean-field calculations of quadrupole deformation parameters for the Fe isotope chain, these results benchmark a maximum of deformation in the $N = 40$ island of inversion in Fe, and shed light on trends in level densities indicated in the newly-refined mass surface.

%\begin{description}
%\pacs{21.10.Dr}
%\end{description}
\end{abstract}

%\pacs{21.10.Dr}% PACS, the Physics and Astronomy
\maketitle

The evolution of nuclear structure away from stability is a primary focus of low-energy nuclear physics. This concerns the change in nuclear stability with proton number, $Z$, and/or neutron number, $N$, departing from the spherical-shape (`magic') shell-closures at 2, 8, 20, 28, 50, 82 and 126. Among the most powerful theoretical approaches are the nuclear mean-field and the spherical shell model. %, the later of which is not to be confused with the same name given in the past to the Nilsson model. %\textcolor{red}{please dont forget the ab-initio calculations, perhaps good to also mention those at the end of the intro part..}. 
Underlying the magic spherical shell closures, we find the single-nucleon energy levels with an ordering, as shown e.g. in Fig.~4.4 of \cite{hornyak1975}. For orbitals between $Z=20$ and $Z=50$, we find the following order: $1f_{7/2}$, $2p_{3/2}$, $1f_{5/2}$, $2p_{1/2}$ and $1g_{9/2}$. 

Within %the mean-field terminology
this terminology, certain variations in the energies of the above order are called inversions. Groups of nuclei for which such a different order applies are referred to as {\em Islands of Inversion}; they are often signaled by an approaching of levels as functions of $Z$ and/or $N$, permitting increased neutron/proton excitations associated with islands of inversion \cite{Ljungvall2010}.

When filling orbitals in between spherical shell closures, we enter an increased level-density zone with shell-energies privileging non-zero quadrupole deformation, a natural degree of freedom in the mean-field approach. Within the spherical shell-model, one expresses the same via an increase of the $B(E2)$, and/or lowering the first excited $I^\pi=2^+$ energies. Both can be seen as differing language for the same physics content of mean-field deformation. The region around $N = 40$ has gained interest due to the presence of such phenomena \cite{Hannawald1999,Sorlin2003,Gade2010,Naimi2012} as potential indicators of deformation and inversion.

\begin{figure}[t]
    \begin{center}
    %\hspace{-10mm}
        \includegraphics[width=0.9\columnwidth]{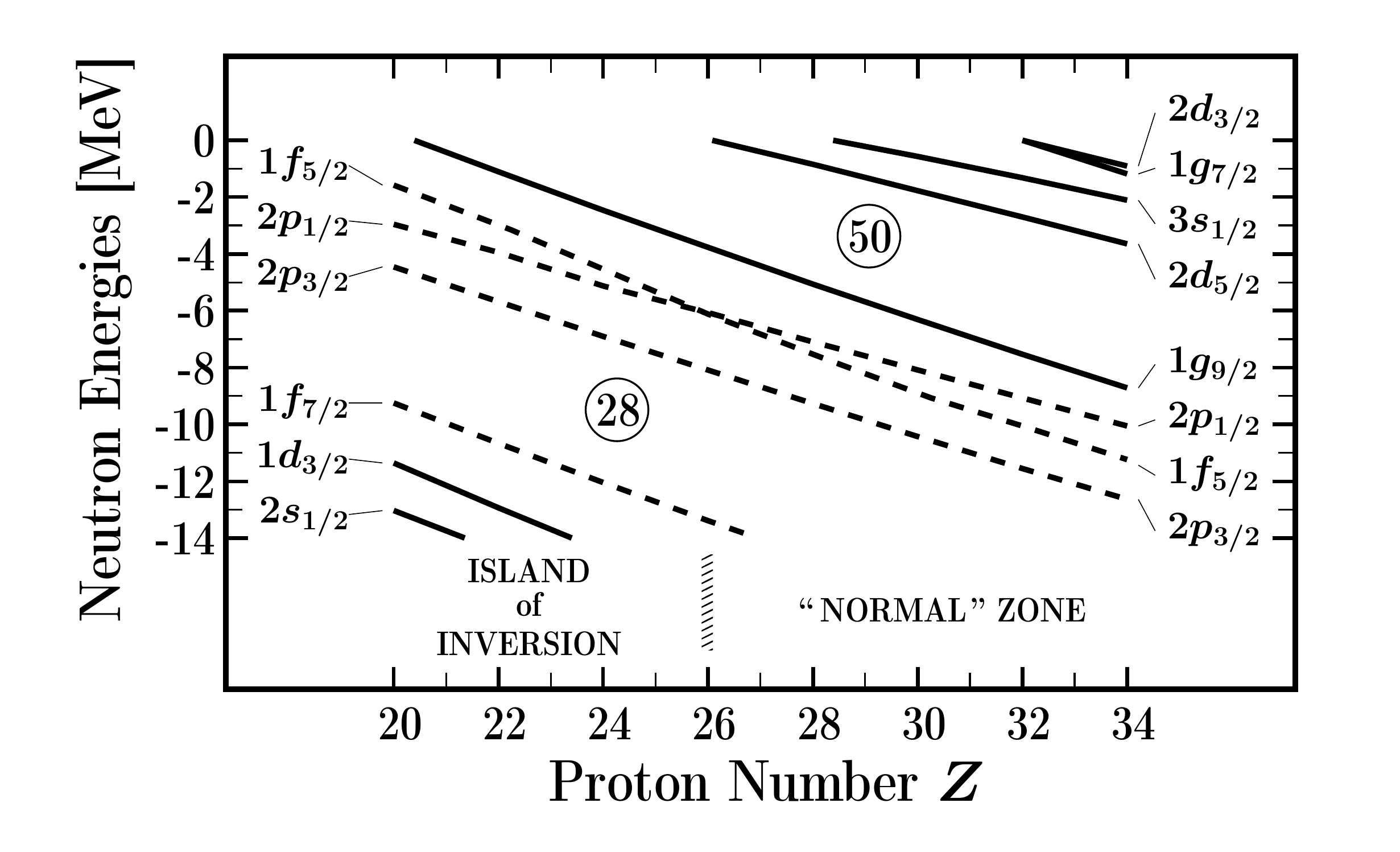}
        \caption{Neutron energies based on the Universal Woods-Saxon Hamiltonian from \cite{Gaamouci2021} along $N = 40$ for spherical shape nuclei; solid (dashed) lines stand for positive (negative) parity orbitals. The close approach of orbitals $2p_{1/2}$ and $1f_{5/2}$ at $Z = 26$ defines a zone of nuclei usually referred to as an ``island of inversion''.
        }
        \label{fig:Island}
    \end{center}
\end{figure}

To examine the single-neutron spectrum in the vicinity of $N = 40$, we have calculated single particle energies as shown in Fig \ref{fig:Island}. Here we follow the state of the art phenomenological realization of the universal mean field approach \cite{Gaamouci2021}, where one set of parameters is common for the whole nuclear chart, parametric correlations of the modeling are analyzed and removed according to the mathematical Inverse Problem Theory. %, and predictive power is tested on the equilibrium deformations for the whole nuclear range of interest. %with no single discrepancy.  
The approach of the $2p_{1/2}$ and $1f_{5/2}$ levels at $Z = 26$ impacts the ground-state structures of nuclei in the two zones, `inversion' and `normal', in different manners, highlighted by the structure differences of the two-body matrix elements of the residual interactions, collective excitation energies and decay strengths. This region has been termed {\em $N=40$ Island of Inversion} in shell-model literature.

Experiments in this neutron-rich $Z = 24-28$ region have been challenging, in large part due to elemental production difficulties. ISOL-type facilities have historically not been considered viable options for the production of transition metals \cite{Kirchner1992,Koster2007}.

Despite experimental challenges, many measurements have played a role in probing the nuclear structure of this region \cite{Block2008,Ferrer2010}. Large $B(E2; 0^+ \to 2^+)$ \cite{Rother2011}, increasing charge radii \cite{Heylen2016} and increasing $E(4^+)/E(2^+)$ ratios \cite{Santamaria2015} have all been interpreted as stemming from the presence of deformed nuclei due to neutron excitations into intruding orbitals, indicative of an island of inversion. On the mass measurement front, previous studies show a flattening behavior in the two-neutron separation energies along the Cr and Mn isotopes \cite{Mougeot2018,Naimi2012}. This is often interpreted as indicative of increasing deformation. %however no explicit confirmation of this connection exists.%
For the Fe isotopes, though data on masses are available, precisions pail in comparison to neighboring isotope chains in the region. Current low resolution experimental data suggests maximum collectivity is reached at $N = 42$ \cite{Meisel2020}, but greater precisions are needed to make definitive conclusions.

In this article, we present precision mass measurements of $^{63-70}$Fe from the TITAN Multiple-Reflection Time-of-Flight Mass Spectrometer (MR-ToF-MS) \cite{Jesch2015}, including the first high-precision ($\delta m/m \sim 10^{-7}$) direct measurements of $^{68-70}$Fe. This represents the first large scale investigation of such neutron-rich Fe at an ISOL facility, supported by resonant laser ion source development. These results are interpreted in terms of mean-field calculations employing recent `universal' Woods-Saxon Hamiltonian \cite{Gaamouci2021} and results from the newly developed multi-shell valence-space in-medium similarity renormalization group (VS-IMSRG) \cite{Miya20lMS}.

\begin{figure}[tb]
    \begin{center}
        \includegraphics[width=0.9\columnwidth]{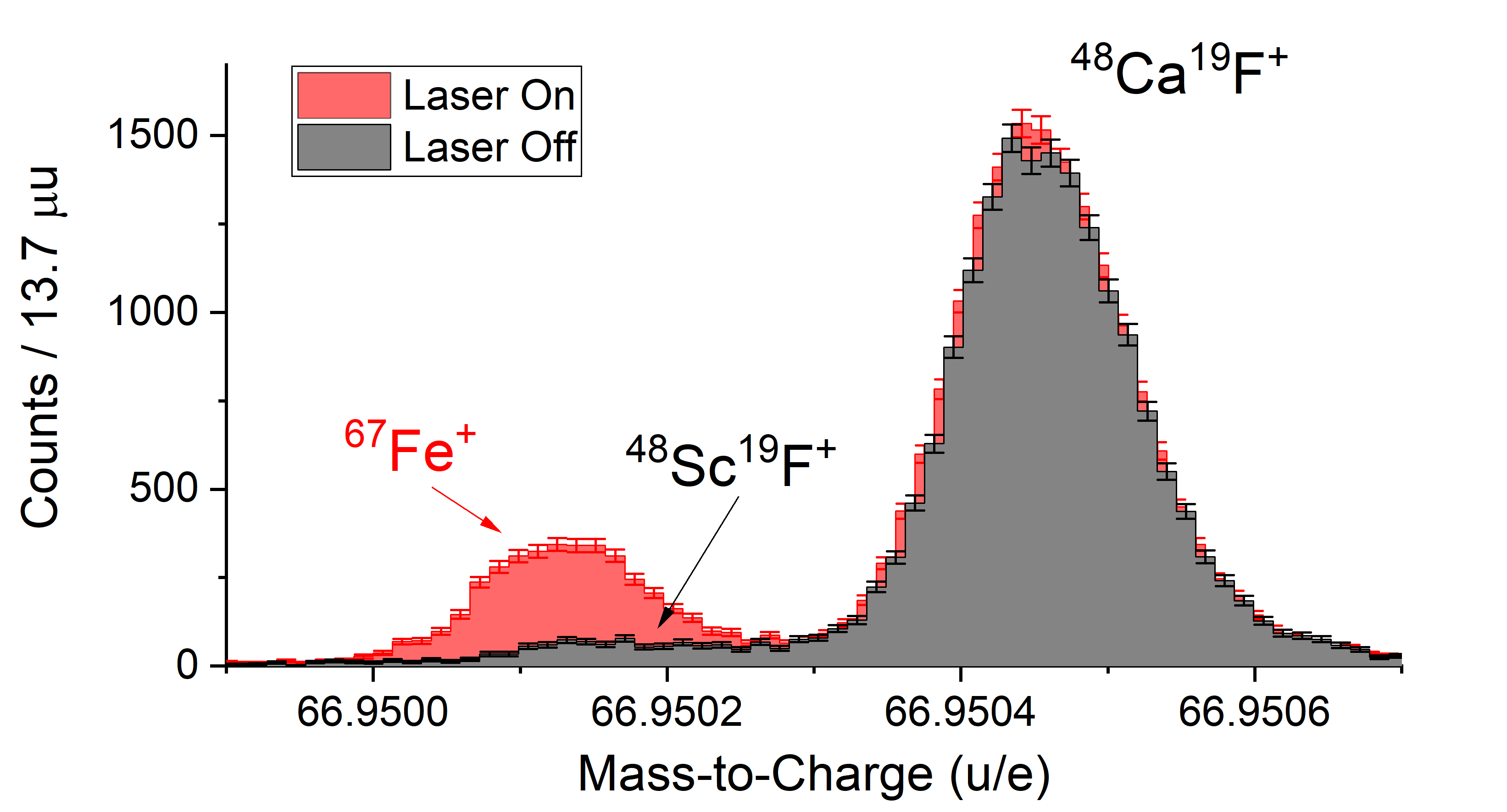}
        \caption{A portion of the mass spectra at $A = 67$ with TRILIS lasers on and off. A factor 4.5 count increase in the leftmost peak with lasers unambiguously confirms its identity as $^{67}$Fe. }
        \label{fig:67Fe_spectra}
   % \end{center}
%\end{figure}
%\begin{figure}[b]
  %  \begin{center}
        \includegraphics[width=0.9\columnwidth]{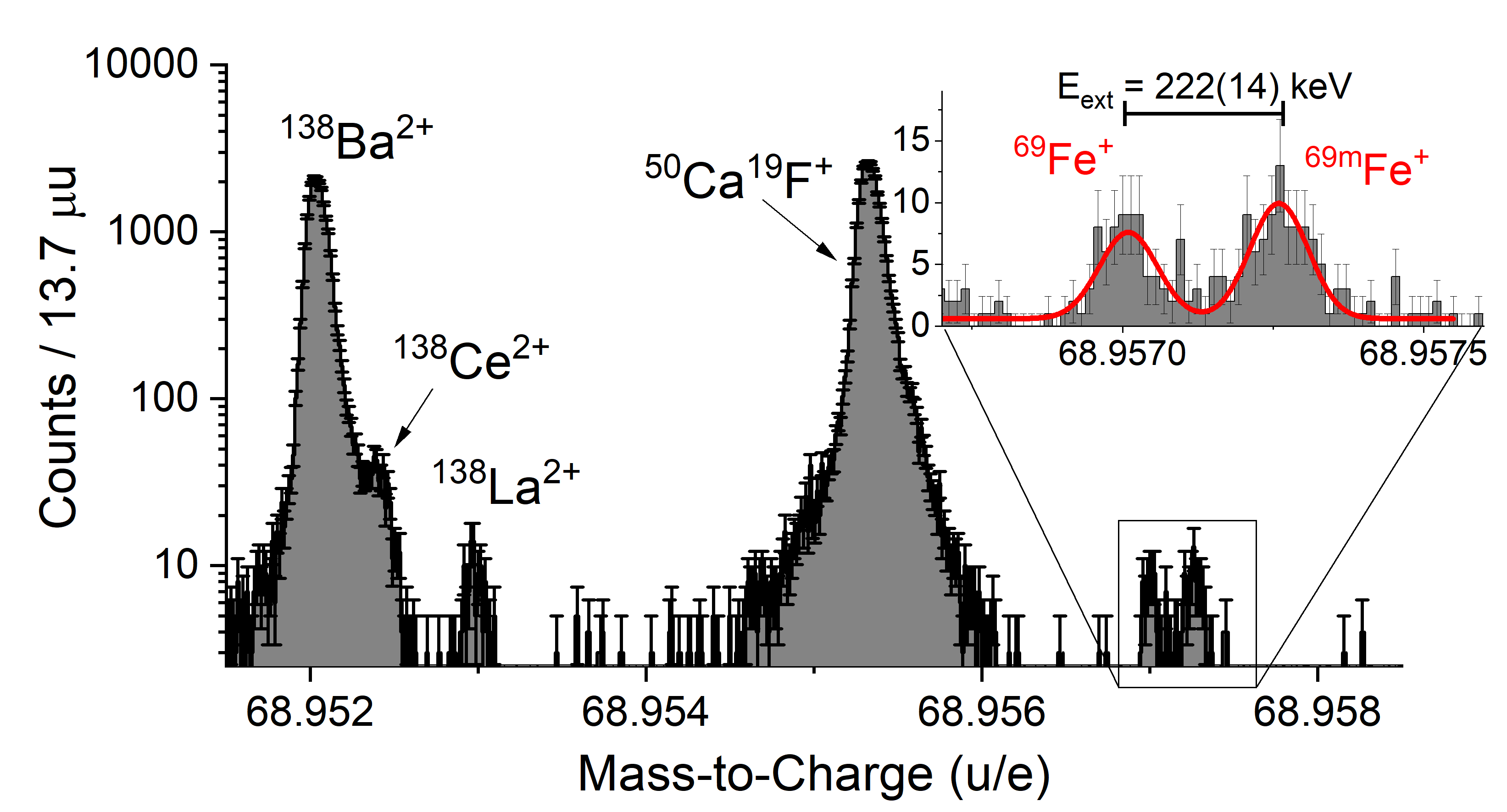}
        \caption{A portion of the mass spectra at $A = 69$. The identified mass species are labeled. %The mass range displayed spans times-of-flight from 6-10 $\mu$s.%
        In the inset, $^{69}$Fe and the newly discovered $^{69m}$Fe are shown.}
        \label{fig:69Fe_spectra}
    \end{center}
\end{figure}

Fe mass measurements were performed at TRIUMF's Ion Trap for Atomic and Nuclear science (TITAN) \cite{Dilling2006}. Fe isotopes were produced at TRIUMF's Isotope Separator and Accelerator (ISAC) \cite{Ball2016}, where a 480 MeV, 20 $\mu$A proton beam was impinged on a UC$_X$ target. Reaction products % as those are n-rich isotopes from an U target, fission actually contributes meaning full.. I simply changed to: reaction products. 
were surface ionized by a hot Re ion source. Further ionization of Fe isotopes was achieved via TRIUMF's resonant ionization laser ion source (TRILIS), using a two-step resonant laser excitation scheme \cite{Lassen2017}. This Fe-specific scheme, used here for the first time, allowed for increased yields by at least two to three orders of magnitude of Fe ions of interest and confirmed the presence of Fe species in spectra, as seen in Fig.~\ref{fig:67Fe_spectra}.

Ionized beams were sent to a mass separator ($R = 2000$) which removed non-isobaric products. The isobaric beam was then transported to the TITAN facility, where it was cooled and bunched via the TITAN radio-frequency quadrupole (RFQ) cooler and buncher \cite{Brunner2012}. Cooled ion bunches were sent to the TITAN MR-ToF-MS \cite{Jesch2015} for mass measurement.

The MR-ToF-MS determines the masses of ions via their time-of-flight over a given path and kinetic energy \cite{Wollnik1990,Plab2013,Reiter2021}. %Since the mass resolution is proportional to the time-of-flight ($R = t/2\Delta t$), a long flight path is desired and achieved via two electrostatic mirrors. Isochronous reflection of ions by these mirrors for a sufficient number of turns achieves the desired resolution \cite{Reiter2021}.% 
After cooling and injection in the system, bunches underwent between 974 to 1008 isochronous turns between electrostatic mirrors for a total time-of-flight of $\sim 16$ ms before ejection onto a MagneToF detector \cite{Stresau2006} for time-of-flight detection. %Such high turn numbers and long total flight time were achievable thanks to first-time operation of the MR-ToF-MS at a lower repitition rate (30 Hz).%
%To minimize contaminant species in our spectra, a mass range selector (MRS) consisting of two electrodes inside the mass analyzer deflected away any remaining non-isobaric beam products \cite{Dickel2015}. % I think the MRS does not add anyhting for the reader... 
To measure and detect masses of very low signal-to-background ratios ($< 1$ to $ 10^{-4}$), %further mass separation is needed%. 
mass-selective re-trapping was employed \cite{Dickel2017Focus,Dickel2017,Beck2021}. %where ions are dynamically recaptured in the injection trap after a defined flight time in the mass analyzer \cite{Dickel2017}. This process is highly mass-selective, and allows for the separation of isobaric contaminants from ions of interest during the mass measurement process \cite{Jacobs2019,Izzo_In2021,Beck2021}. 

%\begin{figure}[b]
 %   \begin{center}
  %      \includegraphics[width=0.93\columnwidth]{69Fe_spectra.png}
   %     \caption{Mass spectra taken with the MR-ToF-MS at $A = 69$. The identified mass species are labeled. %The mass range displayed spans times-of-flight from 6-10 $\mu$s.%
    %    In the inset, $^{69}$Fe and $^{69m}$Fe are shown with linear scale and a fitted hyper-EMG curve in red.}
     %   \label{fig:69Fe_spectra}
    %\end{center}
%\end{figure}

Overall, the measurement process ran with a cycle time of $\sim$ 33 ms, yielding mass resolutions of $R > 600\,000$, the highest achieved with the TITAN MR-ToF-MS. %Such a milestone is due to both first-time operation at a lower repetition rate and additional fine tuning.%
This high resolving powers allowed for separation of a $\sim$ 200 keV isomeric state in $^{69}$Fe as shown in Fig.~\ref{fig:69Fe_spectra}. %For each mass measurement spectra, additional spectra with TRILIS lasers off was taken to unambiguously confirm the identification of Fe species.

Time-of-flight spectra were converted to mass spectra via the relationship: $m_{\text{ion}}/q = C(t_{\text{ion}}-t_0)^2$, where $t_0$ is $0.16$~$\mu$s as determined offline and $C$ is a calibration factor determined from a high-statistics, well-known reference ion in each spectrum as listed in Table \ref{tab:mass_table}. To account for time-dependent drifts, a time-resolved calibration (TRC) was performed \cite{Ayet2019} using the mass data acquisition software package \cite{Dickel2019}. Peak centroids were determined by fitting hyper-EMG functions \cite{Purushothaman2017} using the \verb|emgfit| Python wrapper \cite{emgfit}. Statistical uncertainties are generated based on techniques described in \cite{Ayet2019,emgfit}. Systematic uncertainties of the MR-ToF-MS system are described in detail in \cite{Jacobs2019,Ayet2019}, and total to a value of $\sim 8.6 \times10^{-8}$. This uncertainty is dominated by the uncertainties due to ion-ion interactions ($3.3 \times10^{-8}$ per detected ion), TRCs ($\sim 3.7 \times10^{-8}$), and the non-ideal switching of mirrors ($ 7.0 \times10^{-8}$). For $^{67}$Fe, the presence of underlying counts of $^{48}$Sc$^{19}$F had to be considered, as shown in Fig.~\ref{fig:67Fe_spectra}. Therefore, an uncertainty of 6.7 keV was added in quadrature, following the prescription in \cite{Ayet2019} for the treatment of overlapping peaks. In the case of $^{70}$Fe, where a doubly-charged reference ion was used, an uncertainty of 3.8 keV was added in quadrature to account for differences in peak shape between singly and doubly-charged ions.

Table \ref{tab:mass_table} reports the masses of all measured Fe isotopes, as well their mass excesses as found in literature. The masses of $^{63-66}$Fe have been well-established, and our experimental results are in agreement ($< 0.6\sigma$) with these values. $^{68,69,70}$Fe constitute the first direct mass measurements with precisions of $\delta m/m \sim 10^{-7}$, with previous measurements done via TOF-B$\rho$ mass spectrometry \cite{Meisel2020}. The case of $^{67}$Fe is notable, as our results deviate from a recent Penning trap mass measurement at JYFLTRAP \cite{Canete2020} by $\sim32\sigma$. As seen in Fig.~\ref{fig:67Fe_spectra}, our identification of $^{67}$Fe is confirmed with the species-specific resonant laser excitation scheme via a factor 4.5 count increase of the Fe peak. The mass value reported by \cite{Canete2020} coincides with the $^{48}$Ca$^{19}$F peak, which showed no statistically significant count change with lasers. %$^{48}$Sc$^{19}$F was identified within the Fe peak, and taken into consideration in our analysis. 
Given the expected count increase of Fe species with lasers, our data does not indicate the presence of a long-lived isomer in $^{67}$Fe, which is unexpected given their presence in the other odd-even Fe isotopes. %with the Fe laser excitation scheme. 
In both experiments, the known 387~keV isomeric state in $^{67}$Fe is too short lived ($t_{1/2} = 75 \mu$s) to have been measured with the used techniques.

\begin{table*}[tb]
  \centering
  \caption{Results of the mass measurements performed, compared to the values recommended by AME2020 \cite{Wang2021} for $^{63-67}$Fe and values from \cite{Meisel2020} for $^{68-70}$Fe. Values with a superscript $D$ are unused in the mass calculation in \cite{Wang2021}. We also provide the mass ratio ($m_{\text{ionic,IOI}}/m_{\text{ionic,ref}}$) between the ionic masses of the ion of interest (IOI) and the reference ion. All mass values are in keV. All Fe isotopes were measured as singly-charged ions. Uncertainties presented are total uncertainties calculated as $\sigma_{\text{total}} = \sqrt{\sigma_{\text{sys}}^2 + \sigma_{\text{stat}}^2}$}
    \begin{tabular}{c c c c c c c}
    \toprule    Nuclide & Mass Excess & Literature & Difference & $\quad$ Reference Ion & Mass Ratio \\    \hline

 $^{63}$Fe &  -55635.4(5.4) & -55635.6(4.3) & -0.3\,\,(6.9) &  	$\quad$  $^{44}$Ca\,$^{19}$F\,$^{+}$	 	&  0.999\,792\,4940\,\,(931)	  \\
 
 $^{64}$Fe &  -54970.0(5.3) & -54969.6(5.0) & \,\,0.5\,\,(7.3) &  	$\quad$  $^{48}$Ti\,$^{16}$O\,$^{+}$     &  0.999\,979\,3643\,\,(887)	  \\
 
 $^{65}$Fe &  -51218.7(8.4) & -51217.9(5.1) & \,\,0.8\,\,(9.8) &  	$\quad$  $^{46}$Ti\,$^{19}$F\,$^{+}$	 	&  0.999\,915\,835\,\,\,\,(144)     \\
 
 $^{65m}$Fe &  -50821.4(7.7) & -50823.9(7.2) & -2.5(10.5) &  	$\quad$  $^{46}$Ti\,$^{19}$F\,$^{+}$	 	&  0.999\,922\,402\,\,\,\,(133)	  \\
 
 $^{66}$Fe &  -50061\,\,\,\,\,\,(10) & -50067.8(4.1) & -6.9(11.0) &  	$\quad$  $^{50}$Ti\,$^{16}$O\,$^{+}$	 	&  1.000\,107\,763\,\,\,\,(165)	  \\
 
 $^{67}$Fe &  -46014.8(8.7) & -45708.4(3.8) & 306.3(9.5) &  	$\quad$  $^{48}$Ti\,$^{19}$F\,$^{+}$	 	&  1.000\,071\,788\,\,\,\,(140)	  \\

 $^{68}$Fe &  -44100.9(5.6) & -44360(320)$^D$ & -259\,(320) &  	$\quad$ $^{68}$Ga\,$^{+}$	&  1.000\,371\,3423\,(916)  \\
 
 $^{69}$Fe &  -39504\,\,\,\,\,\,(11) & -40270(400)$^D$ & -767\,(400) &    $\quad$ $^{50}$Ti\,$^{19}$F\,$^{+}$	&  1.000\,216\,864\,\,\,\,(164)	 \\
 
 $^{69m}$Fe & -39281.7(9.1) & -- & --  &  $\quad$ $^{50}$Ti\,$^{19}$F\,$^{+}$		&  1.000\,220\,318\,\,\,\,(142)	 \\ 
 
 $^{70}$Fe & -37053\,\,\,\,\,\,(12) & -37710(490)$^D$ & -657\,(490)  &  $\quad$ $^{140}$Ba\,$^{2+}$	&  1.000\,085\,987\,\,\,\,(213)	 \\ 
 \hline

    \end{tabular}%
  \label{tab:mass_table}%
\end{table*}

Alongside high-precision ground state mass measurements, a long-lived isomeric state in $^{69}$Fe was discovered, and evaluated with excitation energy of $E_x = 222(14)$~keV, see Fig.~\ref{fig:69Fe_spectra}. This identification is strengthened by the observation of the well-known isomeric state in $^{65m}$Fe, whose excitation energy was determined to be $E_x =$ 397(11) keV, in excellent agreement with literature \cite{Kondev2021}. This represents another odd-even isomer in the Fe isotope chain. In order to estimate the half-life of the new isomeric state, a storage time measurement was performed, similar to \cite{Mukul2021}, which suggests a half-life on the order of $100-200$ ms. The detected isomer-ground state pairs for $^{65}$Fe and $^{69}$Fe had isomer-to-ground count ratios of 3.31(3) and 1.33(17), respectively. %, comparable to the half-life of the $^{69}$Fe ground state.
\begin{figure}[t]
    \begin{center}
    \includegraphics[width=0.9\columnwidth]{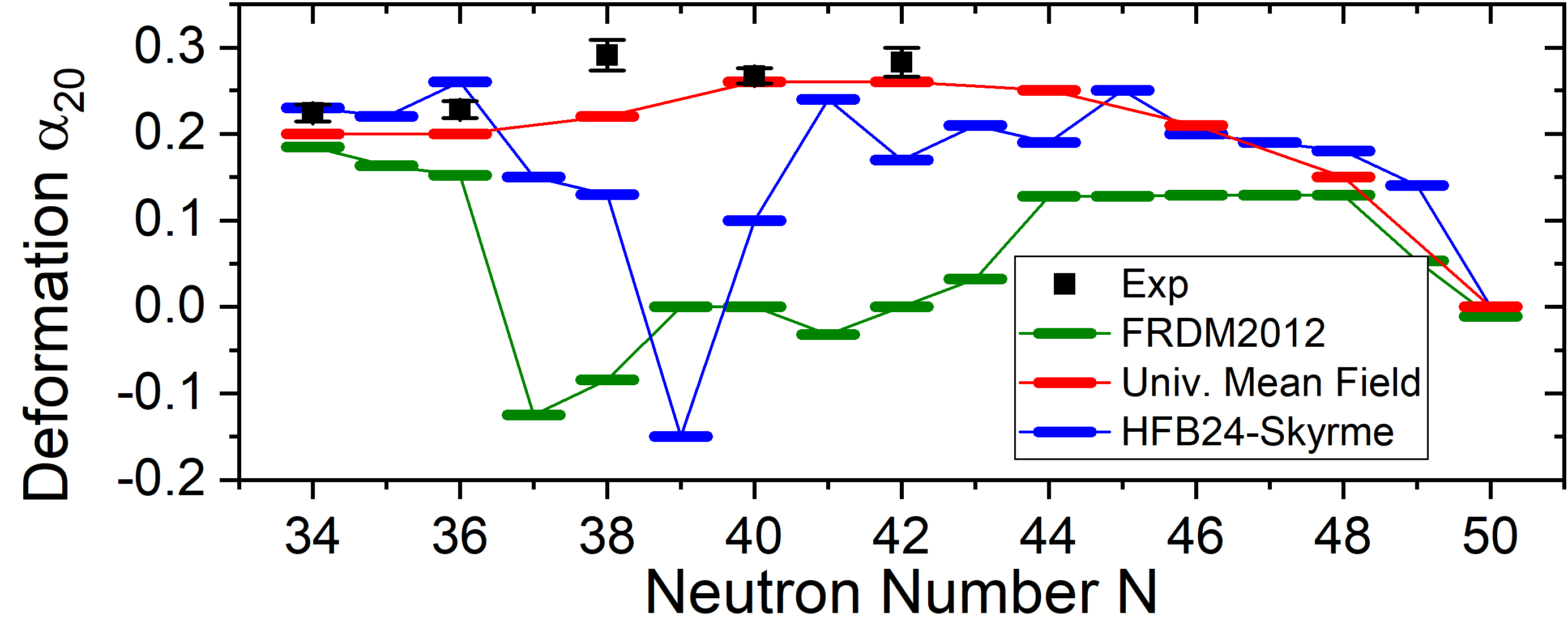}
        \caption{Experimental quadrupole deformation for Fe isotopes from \cite{Pritychenko2016} in comparison to theoretical calculations.}
        \label{fig:quad}
        
  %  \end{center}
%\end{figure}
%\begin{figure}[b]
 %   \begin{center}
       \includegraphics[width=0.900\columnwidth]{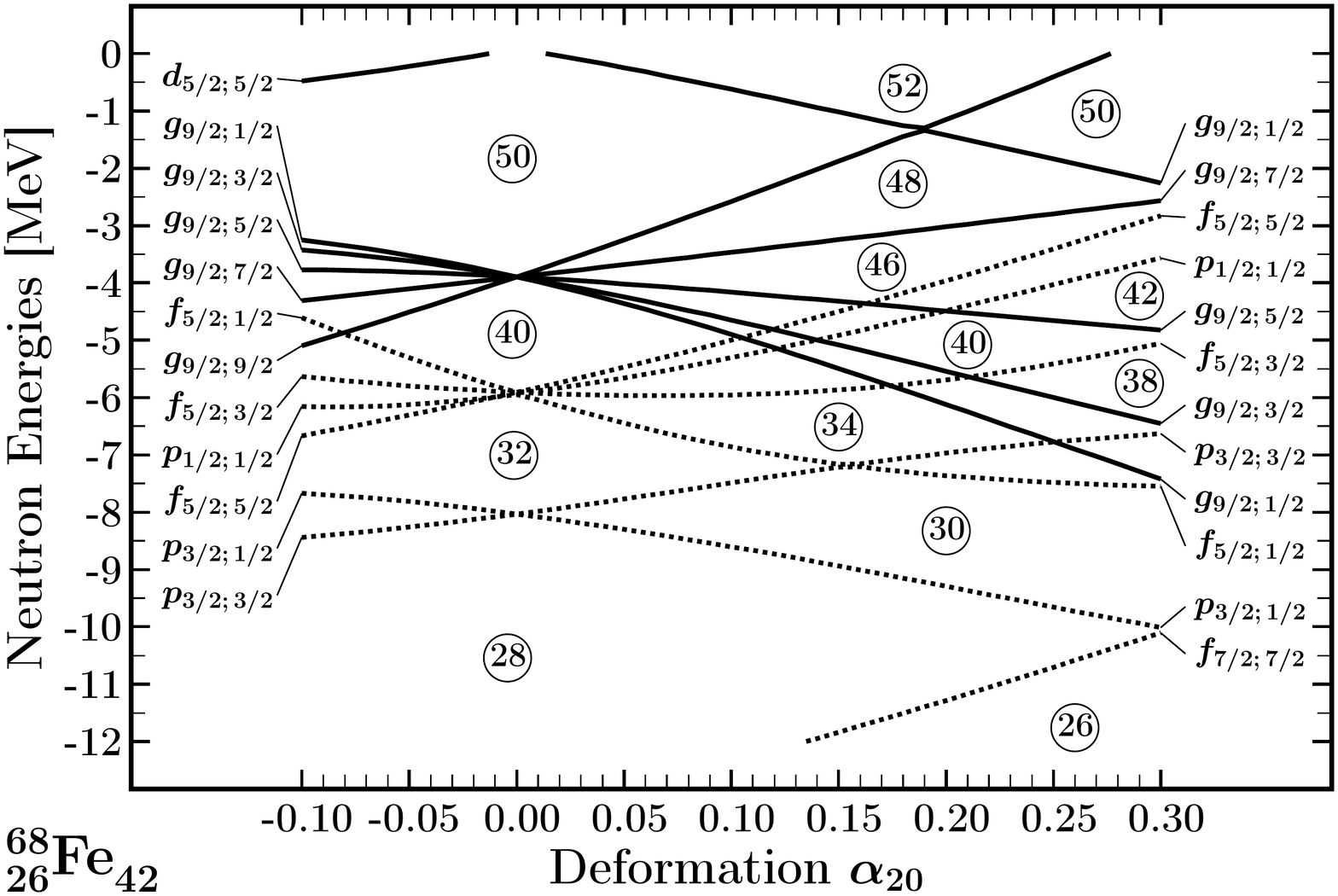}
        \caption{Mean field single-particle neutron energies obtained with the Woods-Saxon Hamiltonian from \cite{Gaamouci2021}, showing neutron orbital patterns in $^{63-69}$Fe nuclei. Note that $p_{1/2}$ and $f_{5/2}$ orbitals, placed between $N = 32$ and $N = 40$ occupations, are nearly degenerate at $\alpha_{20}\approx0$.
         }
        \label{fig:sp-levels}
    \end{center}
\end{figure}
It is not possible to deduce the structure of $^{69}$Fe states from our experiment since the spin-parity assignments are not known. We propose an interpretation based on mean-field calculations and the properties of the sequence of $^{63-69}$Fe isomers, where experimental spin-parity assignments remain inconclusive past $^{65}$Fe. 

Fig \ref{fig:Island} shows the transition into the {\em $N=40$ Island of Inversion}, seen, for example, in the approaching of the $2p_{1/2}$ and $1f_{5/2}$ levels in spherical nuclei. However, known equilibrium shapes across $N=40$ \cite{Pritychenko2016} are all non-spherical and manifest systematically prolate shapes, which will further shift the corresponding energy levels. We performed large-scale mean-field calculations, employing the `universal' Woods-Saxon Hamiltonian \cite{Gaamouci2021}, and extracted deformation and single-particle energy levels. Our calculations are in good agreement with known experimental deformations across $N=40$ and predict a return to sphericity approaching $N=50$, as shown in Fig.~\ref{fig:quad}. The trends at $N=40$ deviate from FRDM \cite{MOLLER20161} and HFB \cite{PhysRevC.88.061302} calculations. In Fig.~\ref{fig:sp-levels} we show the single-particle neutron energies for different quadrupole deformations $\alpha_{20}$.

In $^{65}$Fe, the experimental $I_{\rm g.s.}^\pi = (1/2^-)$ \cite{Kondev2021} corresponds to the 39$^{\rm th}$ neutron occupying the $p_{1/2}$-orbital very close to the $N = 40$ gap at $\alpha_{20}\sim 0.20$,  as in Fig.~\ref{fig:sp-levels}. The $I_{\rm i.s.}^\pi = (9/2^+)$ isomer at 402 keV corresponds to the first excited orbital, $g_{9/2}$, and the nearby $I_{\rm i.s.}^\pi = (5/2^-)$ at 397 keV to $f_{5/2}$.

Experimentally, the 41$^{\rm st}$ neutron in $^{67}$Fe has an uncertain spin-parity assignment, either $5/2^+$ or $7/2^+$ \cite{Kondev2021}, in conflict with $I_{\rm g.s.}^\pi = 1/2^-$ together with $5/2^-$ at 366.4 keV proposed in \cite{Daugas2011}. The latter assignments are in good qualitative correspondence with the closeness of $p_{1/2}$ and $f_{5/2}$ orbitals at $\alpha_{20}\sim 0.20$.

According to our calculations, $^{68}$Fe and $^{70}$Fe are strongly quadrupole deformed, in agreement with the experimental value $\alpha_{20}=0.283$ for $^{68}$Fe \cite{Pritychenko2016}. From the perspective in Fig.~\ref{fig:sp-levels}, at $\alpha_{20}$ exceeding $0.20$, the $43^{\rm rd}$ neutron in $^{69}$Fe occupies the $p_{1/2}$ orbital, with the closely-lying $f_{5/2}$ and $g_{9/2}$ orbitals giving rise to $I^\pi=5/2^-$ and $9/2^+$ excited states. Assuming decays are via $\gamma$-deexcitations given global isomeric properties \cite{Walker2020} and preexisting data in the Fe chain \cite{Kondev2021}, the isomeric candidate $I^\pi_{\rm i.s.} = 9/2^+$ would decay via an $M4$ transition, whereas the $5/2-$ could decay via an $E2$ transition. Given $E_\gamma = 222$ keV, our elementary Weisskopf modeling indicates the $M4$ decay is most likely for a long-lived isomer given the Woods-Saxon diagram expectations and our measured half-life. Thus a tentative spin assignment of $I^\pi_{\rm i.s.} = 9/2^+$ is favored.

To confirm our spin-parity assignment methodology, we checked the $N=41^{\rm st}$ nucleon configuration in $^{69}$Ni, taking into account that, according to our calculations, $^{68}$Ni and $^{70}$Ni are almost spherical. At small deformation, $\alpha_{20} < 0.1$, the 41$^{\rm st}$ neutron occupies the $g_{9/2}$ orbital; experimentally, $I_{\rm g.s.}^\pi = 9/2^+$ is in agreement with our arguments. The lowest lying excited states have $I^\pi = 1/2^-$ and $5/2^-$, again compatible with the near degeneracy of the $p_{1/2}$ and $f_{5/2}$ orbitals in Fig.~\ref{fig:sp-levels}.

\begin{figure}[tb]
    \begin{center}
        \includegraphics[width=0.84\columnwidth]{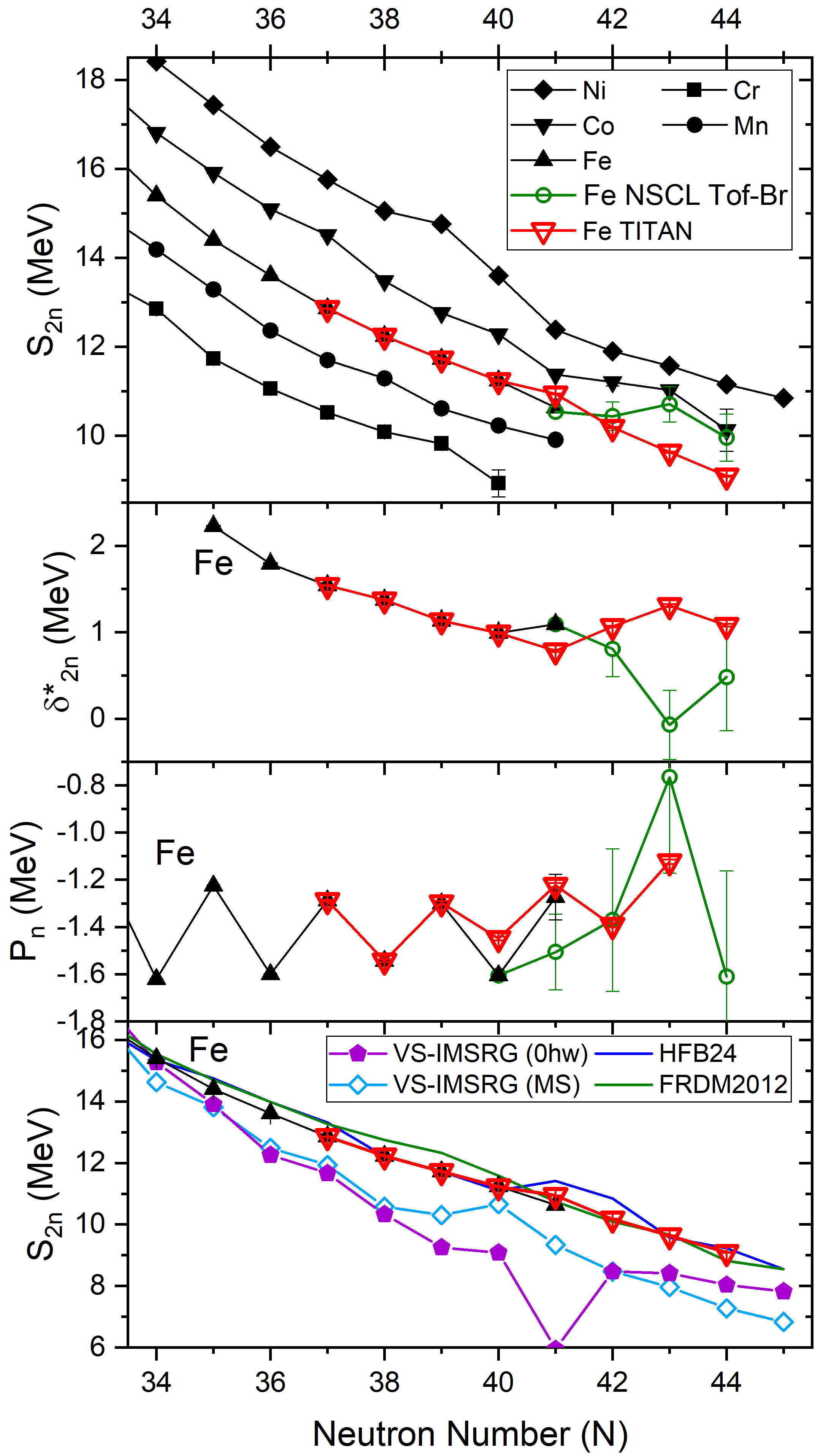}
        \caption{(Top) A graph of the $S_{2n}$ in the $Z = 24-28$ isotope chains based on mass values from \cite{Wang2021} and \cite{Meisel2020}, and $^{63-70}$Fe mass measurements from this work. Of note, the new high-precision TITAN measurements reach as far as the previous ToF-B$\rho$ results from \cite{Meisel2020}. (Middle top) A graph of the $\delta^*_{2n}$ for the Fe isotope chain, showing a minimum at $N = 41$. (Middle bottom) A graph of the pairing gap ($P_n$) from three-point binding-energy differences for the Fe isotopes. (Bottom) A graph of Fe $S_{2n}$ experimental values compared to theoretical calculations for VS-IMSRG results from \cite{Stroberg2021} (0$\hbar\omega$) and this work (MS). The discrepant points at $N = 40$ (MS) and $N = 41$ (0$\hbar\omega$) are shown for completeness, despite the fact they are artifacts of mismatched neutron valence spaces.
        }
        \label{fig:s2n}
    \end{center}
\end{figure}

Fig.~\ref{fig:s2n} shows the two-neutron separation energies ($S_{2n}$) for the $Z = 24 - 28$ isotopes, defined as: $S_{2n}(N,Z) = m(N-2,Z) + 2m_n - m(N,Z)$, with $m(N,Z)$ the mass excess and $m_n$ the mass of the neutron. Our results show an upward curvature in the $S_{2n}$ trend in the Fe isotope chain up until $N = 41$. Also shown is the ($\delta^*_{2n}$), given as: $\delta^*_{2n} = S_{2n}(N-2,Z) - S_{2n}(N,Z)$. As a pseudo-derivative of $S_{2n}$, low $\delta^*_{2n}$ values decreasing towards zero indicate a flattening in $S_{2n}$, whereas a sharp peak indicates a steep drop-off in $S_{2n}$ at $N - 2$. Such sharp peaks, when at even neutron numbers, are typically indicative of bigger energy level gaps \cite{Leistenschneider2018}. The local maximum at $N = 43$, as opposed to $N = 42$, therefore indicates the absence of a large level gap for Fe at $N = 40$. The neutron pairing gap ($P_n$) is also shown, given as: $P_n(N,Z) = (-1)^N[B(N+1,Z) + B(N-1,Z) - 2B(N,Z)]/2$, where $B(N,Z)$ is the binding energy for which we use a positive sign convention. Our high-precision results show a smooth reduction in the odd-even staggering of the pairing gap as $N = 42$ is approached, in contrast to previous results from \cite{Canete2020} and \cite{Meisel2020}.

To further interpret the refined mass surface around $N = 40$, we compare the trends with trends in deformations in the Fe isotope chain, shown in Fig.~\ref{fig:quad}. Our calculations predict the onset of quadrupole equilibrium deformations beginning at $^{60}$Fe with $\alpha_{20}=0.20$, followed by $\alpha_{20} =$ 0.20, 0.22, 0.26, 0.26, 0.25 and 0.21 for $^{62,64,66,68,70,72}$Fe, respectively. The maximum quadrupole deformation occurs between $^{66}$Fe and $^{68}$Fe (i.e. $N = 40$ and $N = 42$). This corresponds with the $\delta^*_{2n}$ minimum at $N = 41$, indicating the maximum level density, and therefore maximum deformation, is correlated to the minimum change in $S_{2n}$. These results reveal the deformation maximum point in the Fe isotope chain. In comparison, here the FRDM \cite{MOLLER20161} and HFB \cite{PhysRevC.88.061302} calculations follow the refined $S_{2n}$ trend with only small fluctuations, despite predicting nearly spherical shapes for the Fe nuclei across $N = 40$.

%\begin{figure}[ht]
%    \begin{center}
%        \includegraphics[width=0.9\columnwidth]{Fe_s2n_models.png}
%        \caption{$S_{2n}$ values in the Fe isotope chains from this work and  leading theoretical models. Empty red squares are from AME2020 \cite{Wang2021} and filled red squares are Fe mass measurements from this work.}
%        \label{fig:s2n_models}
%    \end{center}
%\end{figure}

To examine the results within a different theoretical framework, we present ab initio predictions using VS-IMSRG~\cite{Herg16PR,Stro17ENO,Stro19ARNPS} calculations. We use the 1.8/2.0(EM) NN+3N interaction~\cite{Hebe11fits,Simo17SatFinNuc}, which globally predicts ground-state energies at least to the $^{132}$Sn region ~\cite{Stroberg2021,Morr18Tin,Miya21Heavy}. In particular, separation energies below and above the {\em $N=40$ Island of Inversion} region \cite{Leistenschneider2018,Mougeot2018,Xu19Sc,Leistenschneider2021} have generally been well reproduced. %, 
Following the $0\hbar\omega$ approach in \cite{Stroberg2021}, we take a $pf$ and $sdg$ neutron valence space below and above $N=40$, respectively. This method, however, makes predictions across $N=40$ difficult. To obtain a more reliable description across the $N=40$ gap, we employ the newly developed multi-shell (MS) approach \cite{Miya20lMS}, where we take a neutron $p_{1/2}$, $p_{3/2}$, $f_{5/2}$, $g_{9/2}$ space above a $^{48}$Ca core. The resulting valence-space Hamiltonians were diagonalized with the KSHELL code \cite{KSHELL}. $S_{2n}$ values from the two prescriptions are shown in Fig.~\ref{fig:s2n}. 

Both methods show an offset compared to the experimental $S_{2n}$ beyond $N>35$, also observed in Cr \cite{Mougeot2018}. The behavior is a consequence of the negligence of many-body correlations beyond the VS-IMSRG(2) approximation \cite{Stro17ENO} and implies an enhancement in collectivity beyond $N>35$, in agreement with our mean field calculations. The discrepant point at $N = 40$ ($N=41$ for the $0\hbar\omega$ calculations) is an artifact of continuations due to remaining mismatched $pf$ and $sdg$ neutron valence spaces. Nonetheless, the MS values display similar trends to the experimental mass surface. Though there is still a deviation between our experimental results and the MS calculations, significant improvement is shown over the previous $0\hbar\omega$ VS-IMSRG calculation method. This paves the way towards a complete ab initio picture of the {\em $N=40$ Island of Inversion}.%; currently in development. 

%Further, the VS-IMSRG results predict a $^{69}$Fe isomeric state with either $I^\pi_{\rm i.s.} = 5/2^-$ or $1/2^-$. In $^{65}$Fe, results indicate a spin assignment of $I^\pi_{\rm i.s.} = 9/2^+$.  These assignments are in agreement with the mean-field theoretical assignments. The change in parity of isomeric states in Fe from positive to negative is validated by two fundamentally different and universal theoretical approaches, both supporting the evolution of orbitals in a similar fashion.

In conclusion, we have presented the measurements of neutron-rich Fe isotopes performed at the TITAN facility using its MR-ToF-MS. These results represent a significant increase in precision for three first-time direct mass measurements of a precision $\sim 10^{-7}$, and include the discovery of a new long-lived odd-even isomer, accomplished with the first-time use of Fe beams at an ISOL facility. The pairing of the highly-efficient TITAN MR-ToF-MS system with an Fe specific laser ionization scheme enabled the reach of exotic Fe isotopes comparable to that of the NSCL ToF-B$\rho$ measurements, paving the way for future high-precision mass campaigns. These measurements, together with the pairing of coherent interpretations employing state-of-the-art mean-field and ab initio calculations, refine our understanding of nuclear structure around the {\em $N=40$ Island of Inversion}, and establish a deformation maximum point in the Fe isotope chain. Further spectroscopic data are needed to experimentally determine the spin of the newly discovered $^{69}$Fe isomer. Nonetheless, the connection of the mass surface to nuclear deformation %shown for the first time, is a major leap forward,%
showcases the importance of nuclear masses in the search for deformation across all regions of the nuclear chart. In particular, more neutron-rich precision mass measurements in the $Z = 24-28$ region are needed to resolve nuclear structure effects as $N = 50$ is approached. 

\begin{acknowledgments}

The authors would like to thank S. R. Stroberg for the imsrg++ code~\cite{Stro17imsrg++} used to perform VS-IMSRG calculations and J. Lassen and the laser ion source group at TRIUMF for their development of the Fe laser scheme. This work was supported by the Natural Sciences and Engineering Research Council (NSERC) of Canada under grants SAPIN-2018-00027 and RGPAS-2018-522453, and by the National Research Council (NRC) of Canada through TRIUMF, the Polish National Science Centre under Contract No. 2016/21/B/ST2/01227, the Polish-French COPIN-IN2P3 collaboration agreement under project numbers 04-113 and 05-119 and COPIGAL 2020, the UKRI Science and Technology Facilities Council (STFC) grant No. ST/P004008/1, the U.S. Department of Energy, Office of Science, Office of Nuclear Physics under grant DE-FG02-93ER40789, and the German Research Foundation (DFG), grant No. SCHE 1969/2-1, by the German Federal Ministry for Education and Research (BMBF), grant No. 05P19RGFN1 and 05P21RGFN1, by the Hessian Ministry for Science and Art through the LOEWE Center HICforFAIR, by the JLU and GSI under the JLU-GSI strategic Helmholtz partnership agreement. Support from the National Natural Science Foundation of China, grant No. 11975209, and the Physics Research and Development Program of Zhengzhou University, grant No. 32410017 is acknowledged. Computations were performed with an allocation of computing resources on Cedar at WestGrid and Compute Canada, and on the Oak Cluster at TRIUMF managed by the University of British Columbia department of Advanced Research Computing (ARC).

\end{acknowledgments}

%\begin{thebibliography}
\bibliography{library}
%\end{thebibliography}
    
\end{document}